\documentclass[journal,draftcls,onecolumn,12pt,twoside]{IEEEtran}
\normalsize

\usepackage{ifpdf}
\usepackage{amsmath,epsfig}
\usepackage{amssymb, amsfonts}
\usepackage{amsthm}
\usepackage[caption=false,font=footnotesize]{subfig}
\usepackage{paralist}
\usepackage{bm}
\usepackage{cite}
\usepackage{color}
\usepackage{epstopdf}
\usepackage{multirow}

\usepackage{dsfont}

\usepackage{tikz}
\usetikzlibrary{arrows,shapes,backgrounds,plotmarks,positioning}
\usepackage{pgfplots}
\pgfplotsset{grid style={dashed,gray}}

\usepackage{pgfplotstable}
\usepackage{booktabs}
\usepackage{colortbl}
\usepackage{multirow}
\usepackage{url}

\theoremstyle{theorem}
\newtheorem{theorem}{Theorem}
\newtheorem{lemma}[theorem]{Lemma}

\newtheorem{proposition}[theorem]{Proposition}

\theoremstyle{definition}
\newtheorem{definition}{Definition}

\theoremstyle{remark}
\newtheorem{remark}{Remark}

\newcommand{\packet}{\mathbf{b}}
\newcommand{\chunk}{\mathbf{B}}
\newcommand{\chunkset}{\mathcal{I}}
\newcommand{\rank}{\mathrm{rk}}
\newcommand{\ex}{\mathrm{E}}

\begin{document}
%
\title{Expander Chunked Codes}
%
%
%
\author{Bin~Tang,
        Shenghao~Yang,~\IEEEmembership{Member,~IEEE,}
        Baoliu~Ye,~\IEEEmembership{Member,~IEEE,}
        Yitong~Yin,~and~Sanglu~Lu,~\IEEEmembership{Member,~IEEE}
\thanks{This paper was presented in part at the IEEE International Symposium on Information Theory, Cambridge, MA, United States,
  July 1-6,  2012.}
\thanks{B. Tang, B. Ye, Y. Yin, and S. Lu are with the National Key Laboratory for Novel Software Technology, Nanjing University, Nanjing, China (e-mail: \{tb, yebl, yinyt, sanglu\}@nju.edu.cn).}
\thanks{S. Yang is with the Institute of Network Coding, The Chinese
  University of Hong Kong, Hong Kong, China (e-mail: shyang@inc.cuhk.edu.hk).}}

\maketitle

\begin{abstract}
Chunked codes are efficient random linear network coding (RLNC) schemes with low computational cost, where the input packets are encoded into small chunks (i.e., subsets of the coded packets). During the network transmission, RLNC is performed within each chunk. In this paper, we first introduce a simple transfer matrix model to characterize the transmission of chunks, and derive some basic properties of the model to facilitate the performance analysis. We then focus on the design of overlapped chunked codes, a class of chunked codes whose chunks are non-disjoint subsets of input packets, which are of special interest since they can be encoded with negligible computational cost and in a causal fashion.
We propose expander chunked (EC) codes, the first class of overlapped chunked codes that have an analyzable performance,
 where the construction of the chunks makes use of regular graphs. Numerical and simulation results show that in some practical settings, EC codes can  achieve rates within 91 to 97 percent of the optimum and outperform the state-of-the-art overlapped chunked codes significantly.
\end{abstract}

%
\IEEEpeerreviewmaketitle

\section{Introduction}

Random linear network coding (RLNC) has great potential for data dissemination over communication networks \cite{AhlswedeTIT, LiTIT, HoISIT, HoTIT}. RLNC can be implemented in a distributed fashion due to its random nature, and is shown to be asymptotically capacity-achieving for networks with packet loss in a wide range of scenarios~\cite{WuISIT, Dana2006, LunPC}. In this paper, we propose a low complexity RLNC scheme called \emph{Expander Chunked (EC)} codes and analyze the achievable rates of EC codes.

\subsection{Background}

For ordinary RLNC studied in literature \cite{HoISIT, HoTIT,WuISIT, Dana2006, LunPC}, all participating nodes forward coded packets formed by random linear combinations of all the packets received so far. Major issues in applying ordinary RLNC include the computational cost and the coefficient vector overhead. Consider the dissemination of $k$ input packets, each consisting of $L$ symbols from a finite field. For encoding, RLNC requires $\mathcal{O}(kL)$ finite field operations to generate a coded packet, and for decoding, a destination node takes $\mathcal{O}(k^2+kL)$ finite field operations per packet if Gaussian elimination is employed. Moreover, to recover the transfer matrices of network coding at the destination node, a coefficient vector of $k$ symbols is usually included in each of the transmitted packets \cite{HoISIT}. Since the packet length $L$ has an upper bound in real-world communication networks,\footnote{For example, network protocols usually have a maximum transmission unit (MTU) ranging from hundreds to thousands bytes.} using large values of $k$ reduces the transmission efficiency.  When there are hundreds of input packets, the computational cost and the coefficient vector overhead would make RLNC difficult for real-world implementation.

To resolve these issues, \emph{chunked (network) codes} have been proposed~\cite{ChouAllerton}, where the input packets are encoded into multiple small \emph{chunks} (also called generations, classes, etc.), each of which is a subset of the coded packets. When using chunked codes, an intermediate network node can only combine the packets of the same chunk.  The encoding and decoding complexities per packet of chunked codes are usually $\mathcal{O}(mL)$ and $\mathcal{O}(mL+m^2)$, respectively, where $m$ is the \emph{chunk size}, i.e., the number of packets in each chunk. The coefficient vector overhead also reduces to $m$ symbols per packet since only the transfer matrices of the chunks are required at the destination nodes. Even so, the chunk size should be a small value (e.g., 16 or 32) for the purpose of practical implementation, as demonstrated in~\cite{LiuINFOCOM}.


Existing chunked codes are in two categories: \emph{overlapped chunked codes} and \emph{coded chunked codes}. In overlapped chunked codes, the chunks are subsets of the input packets with possibly non-empty intersections. The first several designs of chunked codes all belong to this category. However, the existing designs of overlapped chunks are mostly based on heuristics, and no rigorous performance analysis is available for the existing designs~\cite{SilvaNetCod, HeidarzadehITW, LiTIT2011}. In coded chunked codes, chunks are generated by combining multiple input packets. By generalizing fountain codes and LDPC codes, nearly throughput optimal chunked codes have been designed, including BATS code \cite{yang11ac,yang14} and Gamma code \cite{MahdavianiNetCod,MahdavianiNetCod13}. 
Overlapped chunks can be viewed as a degraded class of coded chunks where chunks are generated using certain repetition codes.

Overlapped chunked codes, however, can have lower encoding complexity and latency than general coded chunked codes.  First, as no new packets are necessarily generated during the encoding, the encoding complexity is dominated by generating the indices for the packets in each chunk, which does not depend on the packet length $L$. In contrast, coded chunked codes incur a computational cost that is linear of $L$ to generate a coded packet.
 Therefore, compared to general coded chunked codes, the computational cost of overlapped chunked codes is usually negligible.

Second, overlapped chunks can be encoded in a \emph{causal fashion}. Suppose that the input packets arrive at the encoder gradually. The first chunk can be generated after collecting $m$ input packets, and for every $m$ input packets collected in the following, at least one new chunk can be formed. Therefore, the generation as well as the transmission of chunks can be performed in parallel with the collection of the input packets, reducing the total transmission latency. In contrast, how to achieve causal encoding for general coded chunked codes is not clear: BATS codes and Gamma codes usually require a large fraction of the input packets for encoding chunks.

These advantages motivate us to study overlapped chunked codes, which are especially suitable for delay sensitive applications and networks where the source node has limited computation and storage power, e.g., wireless sensors and satellites.

\subsection{Our Contribution}

We propose \emph{expander chunked (EC) codes}, the first class of overlapped chunked codes that has analyzable performance.  In an EC code, the overlapping between chunks are generated using a regular graph: Each chunk corresponds to a node in the graph and two adjacent chunks share an input packet. EC codes can be encoded causally and share the same belief propagation (BP) decoding of general overlapped chunked codes.

We analyze the BP decoding performance of EC codes generated based on random regular graphs. By exploring the locally tree-like property of random regular graphs and then conducting a tree-based analysis similar to  that of LT/LDPC code,
we obtain a lower bound on the achievable rate depending only on the chunk size, the degree of the regular graph and the rank distribution of the transfer matrices.

The achievable rates of EC codes are evaluated and compared with other chunked codes in two scenarios. We first compare the achievable rates of EC codes with representative coded chunked codes for randomly sampled rank distributions of the transfer matrices, where the purpose is to understand the general performance of EC codes. We find that the performance of EC codes highly depends on the rank distributions: when the expected rank is relatively large, the average achievable rate (over the rank distributions sampled) of EC codes is close to $90\%$ of the representative coded chunked codes, as well as a theoretical upper bound. But for relatively small expected ranks, the achievable rate of EC codes varies significantly for different rank distributions.

To further see the real-world potential of EC codes, we evaluate the performance for a near-optimal chunk transmission scheme over line-topology (line) networks~\cite{bin14}. Line topology itself is of many practical uses, and the scheme for line networks can be extended to general unicast networks and some multicast networks while perserving the performance~\cite{bin14,yang14}. For a wide range of the packet loss rates, with proper optimization of the transmission scheme, EC codes achieve rates very close to those of the coded chunked codes, and about $91\% \sim 97\%$ of the theoretical upper bounds. Besides, we show by simulation that EC codes perform much better than the existing overlapped chunked codes in line networks.



As another contribution, a simple transfer matrix model is proposed to characterize the transmission of chunks over networks with packet loss. Compared with a similar model proposed in \cite{yang14}, which is more suitable for BATS codes, our model incorporates some more practical features of network operations for general chunked codes, making the design of efficient network transmission protocols easier. Therefore, our model is of independent interest for chunked codes. We derive some properties of this transfer matrix model for the performance analysis, which can apply for general chunked codes.



\subsection{Related Work}

The simplest way to form a chunked code is to use disjoint subsets of the input packets as chunks \cite{ChouAllerton}, which has been used in some applications of RLNC~\cite{ChachulskiSIGCOMM, LinICNP,LiuINFOCOM}. To decode a chunk, the transfer matrix of the chunk must have full rank of $m$; otherwise, none of the packets in the chunk could be recovered with high probability. But it is not always a simple task to guarantee the success of decoding a chunk at the destination node. One approach is to use feedback-based chunk transmission mechanism~\cite{ChachulskiSIGCOMM}. While some efficient feedback protocols for specific applications have been developed~\cite{LinICNP,KoutsonikolasTON}, in general, such feedback incurs an inevitable delay and also consumes network resources, resulting in degraded system performance. Besides, for some scenarios such as satellite and deep-space communications, feedbacks are not even available. Another approach is to employ a random scheduling based chunk transmission scheme~\cite{MaymounkovAllerton}, where every network node always randomly selects a chunk for transmission. But this scheme has poor performance for small chunk sizes~\cite{SilvaNetCod, HeidarzadehITW}.

Instead of using disjoint chunks of input packets, chunks with overlaps, i.e., different chunks share some input packets in common, have been proposed by several groups independently~\cite{SilvaNetCod, HeidarzadehITW, LiTIT2011}. It is shown via simulations that overlapped chunked codes have much better performance than disjoint chunks~\cite{SilvaNetCod, HeidarzadehITW}. 
The random annex codes proposed by Li~\textit{et al.}~\cite{LiTIT2011} demonstrate better performance in simulation than the overlapped chunked codes in~\cite{SilvaNetCod, HeidarzadehITW}, but only heuristic analysis of the design is provided.

 BATS code \cite{yang14,yang11ac} is the first class of chunked codes that uses coded chunks. Each chunk in a BATS code is generated as linear combinations of a random subset of the input packets. BATS codes can be regarded as a matrix generalization of fountain codes \cite{lubyLT,Shokrollahi}, and preserve the ratelessness of fountain codes.

Another kind of coded chunked codes consists of chunks that satisfy some parity-check constraints, similar to those of LDPC codes. The first class of such codes is Gamma codes~\cite{MahdavianiNetCod,MahdavianiNetCod13,mahdaviani13}, where the parity-check constraints are applied on the whole chunk~\cite{MahdavianiNetCod}, or on the individual packets in chunks\cite{mahdaviani13}. Another class of such codes is L-chunked codes~\cite{yang14ldpc} which consider more general parity-check constraints and show better performance. Note that the original Gamma codes~\cite{MahdavianiNetCod} paper is published in parallel with the conference version of this paper \cite{tang12}, while the refined Gamma codes~\cite{mahdaviani13} and L-chunked codes are published later than that of our conference version.

Various chunked code based transmission schemes have been designed and implemented recently \cite{huang14mobihoc, yang14a, bin14}, which are consistent with our transfer matrix model.

\section{Overlapped Chunked Codes}
\label{Section: Model}

In this section, we give a general formulation of overlapped chunked codes, including causal encoding and belief propagation (BP) decoding. We also provide a transfer matrix model for general chunked codes.

\subsection{Encoding of Chunks}
\label{sec:encoding-chunks}

Consider transmitting a set of $k$ input packets $\packet_1$, $\packet_2$, $\ldots$, $\packet_k$ from a source node to a destination node over a network with packet loss. Each input packet composes of $L$ symbols from the finite field $\mathbb{F}_q$ with size $q$, and is regarded as a column vector in $\mathbb{F}_q^L$ henceforth.

\begin{definition}[Chunked Codes]
A \emph{chunk} is a set of packets each of which is a linear combination of the input packets, and a \emph{chunked code} is a collection of chunks. A chunked code is said to be \emph{overlapped} if its chunks are subsets of the input packets with possibly non-empty overlapping.
\end{definition}

In this paper, we focus on the design of overlapped chunked codes. Evidently, an overlapped chunked code can be generated by repeating some input packets. Same as most related works, we assume that all the chunks in a chunked code have the same cardinality $m$, which is called the \emph{chunk size}. As the chunk size is related to the encoding/decoding computational complexities and the coefficient vector overhead, for the sake of the applicability in common networks, we regard the chunk size $m$ as a fixed constant which does not change with the number of input packets.

An overlapped chunked code can be more concisely repsesented by a collection of index sets of size $m$. For any integer $n$, let $\chunkset_1,\chunkset_2,\ldots, \chunkset_n$ be subsets of $\{1,\ldots,k\}$ with size $m$.
Let $\chunk_j =\{\packet_i,i\in \chunkset_j\}$. We call either $\chunkset_j$ or $\chunk_j$ a chunk, and the subscript $j$ the chunk ID. An overlapped chunked code of $n$ chunks can be given by either $\{\chunkset_j,j=1,\ldots,n\}$ or $\{\chunk_j, j=1,\ldots,n\}$.

Since each chunk is a subset of the input packets, it is not necessary to duplicate the existing input packets for chunk encoding. During the encoding, only the address in the memory of each packet in a chunk needs to be recorded.

Furthermore, chunks can be encoded \emph{causally} when the input packets arrive at the source node sequentially. By applying proper permutations of the indices, we can always have that the maximum indices among the first $j$ chunks $\cup_{i=1}^j \chunkset_i$ is less than or equal to $mj$. In other words, for any $m$ packets received consecutively, at least one new chunk can be encoded. In this way, the encoding as well as the transmission of chunks can be performed in parallel with the collection of the input packets, so that the total transmission latency can be reduced.

\subsection{Transmission of Chunks}
\label{sec: transmissionmodel}

Each transmitted packet in the network is of the form $(j, \mathbf{c}, \packet)$, where $j$ specifies a chunk ID,
$\mathbf{c}\in \mathbb{F}_q^m$ is the coefficient vector, and $\packet=\chunk_j\mathbf{c}$, a linear combination of packets in $\chunk_j$, is the payload. Here, with some abuse of notation, $\chunk_j$ is also treated as a matrix formed by juxtaposing the packets in $\chunk_j$.  For convenience, we refer to a packet with chunk ID $j$ as a $j$-packet.

Now we describe a chunk transmission model through a network employing linear network coding, which is consistent with the recent design and implementation of
chunked code based network protocols~\cite{huang14mobihoc,yang14a,bin14}.
Consider the $j$-th chunk of packets $\packet_{j_1}$, $\packet_{j_2}$, $\ldots, \packet_{j_m}$. The source node first attaches a coefficient vector to each packet and generates $\tilde \packet_{j_i} = (\mathbf{e}_i, \packet_{j_i})$, $i=1,\ldots,m$, where $\mathbf{e}_i$ is the $i$-th column of the $m\times m$ identity matrix. The source node then generates $M_j$ random linear combinations of $\tilde \packet_{j_i}$ and transmits these linear combinations after attaching the chunk ID, where $M_j$ is an integer-valued random variable.

At an intermediate network node, suppose that $h$ $j$-packets have been received, denoted by $(j, \mathbf{c}^i, \packet^i)$, $i=1,\ldots, h$. The network node can transmit $j$-packet $(j, \mathbf{c}, \packet)$ generated by
 \begin{equation}
   \label{Eq: codingscheme}
    \mathbf{c}=\sum_{i=1}^h\phi_i\mathbf{c}^i,\text{   and   }\packet=\sum_{i=1}^h\phi_i\packet^i,
  \end{equation}
where $\phi_i$, $i=1,2,\ldots,h$ are chosen from $\mathbb{F}_q$. A network node does not transmit combinations of packets of different IDs. Note that in \eqref{Eq: codingscheme}, we only need to combine the $j$-packets with linearly independent coefficient vectors.

At the destination node, let $\mathbf{T}_j$ be the matrix formed by the coefficient vectors of all the $j$-packets received, and
 let $\mathbf{Y}_j$ be the matrix formed by the payloads of all the $j$-packets received. We have
\begin{equation}\label{eq:decoding}
  \mathbf{Y}_j=\chunk_j\mathbf{T}_j,
\end{equation}
where $\mathbf{T}_j$ is called the \emph{transfer matrix} of $\chunk_j$.
Without affecting the decoding performance, we can remove some received $j$-packets so that the remaining set of $j$-packets have linearly independent coefficient vectors. So we assume that $\mathbf{T}_j$ has full-column rank. According to the transmission scheme we describe, we can further write
\begin{equation*}
  \mathbf{T}_j = \mathbf{S}_j\mathbf{H}_j
\end{equation*}
where $\mathbf{S}_j$ is an $m\times M_j$ random matrix corresponding to the linear combinations performed by the source node, and $\mathbf{H}_j$ is a random matrix with $M_j$ rows corresponding to the linear operations performed by intermediate nodes as well as the random packet losses over the network links.
Here for a given value of $M_j$, $\mathbf{S}_j$ is a \emph{totally random} matrix, i.e., every entry of $\mathbf{S}_j$ is chosen from $\mathbb{F}_q$ uniformly and independently at random. Also, we assume that $\mathbf{H}_j$ and $\mathbf{S}_j$ are independent conditioning on $M_j$ and $\rank(\mathbf{S}_j)$, which holds for all the recent chunked code based network protocols~\cite{huang14mobihoc,yang14a,bin14}.

A key result about the transfer matrices is that the column space of each transfer matrix with a fixed dimension is uniformly distributed over all the subspaces with the same dimension. Formally,
\begin{lemma}
\label{lemma: symmetric}
  For any two subspaces $\mathsf{W}$, $\mathsf{U}$ of $\mathbb{F}_q^m$ with the same dimension,
  \begin{equation*}
  \Pr\{\langle\mathbf{T}_j\rangle=\mathsf{W}\}=\Pr\{\langle\mathbf{T}_j\rangle=\mathsf{U}\},
  \end{equation*}
  where  $\langle\mathbf{T}_j\rangle$ denotes the column space of matrix $\mathbf{T}_j$.
\end{lemma}
\begin{IEEEproof}
  See Appendix~\ref{Appendix: transfer matrix}.
\end{IEEEproof}

Assume that $\rank(\mathbf{T}_j)$ follows the probability distribution $t = (t_0,t_1,\ldots,t_m)$, i.e., $\Pr\{\rank(\mathbf{T}_j)=i\}=t_i$ for $i=0,1,\ldots,m$. We further have the following theorem, which is the footstone for the analysis of BP decoding to be described later.
\begin{theorem}
\label{lemma: iterative decoding}
  Let $\mathbf{D}$ be a fixed matrix with $m$ rows and $\rank(\mathbf{D})=w$. Then,
  \begin{equation*}
  \Pr\{\rank([\mathbf{T}_j\text{ }\mathbf{D}])=m\}=\sum_{i=m-w}^m\frac{q^{(m-i)(m-w)}{w \brack m-i}}{{m \brack i}}t_i\triangleq \beta_w,
  \end{equation*}
  where ${w\brack i}=\prod_{j=0}^{i-1}\frac{q^w-q^j}{q^i-q^j}$ is the Gaussian binomial.
\end{theorem}
\begin{IEEEproof}
  See Appendix~\ref{sec:iterativedecoding}.
\end{IEEEproof}

Our chunk transmission model does not depend on a particular chunked code, and hence can be used for the analysis of other chunked codes. A similar model has been used for BATS codes \cite{yang14}. Our model, however, explicitly incorporates a parameter $M_j$ indicating the number of packets transmitted of a chunk, which has a clear operation meaning in chunked code based network protocols. Intuitively, when the network has a higher packet loss rate, we intend to use a larger value of $M_j$ to gain the benefit of network coding. Readers can find more discussion about this parameter in \cite{bin14}.

\subsection{BP Decoding}

The destination node tries to decode the input packets by solving the local linear systems $\mathbf{Y}_j=\chunk_j\mathbf{T}_j$, $j=1,2,\ldots,n$. These local linear systems for chunks jointly give a global linear system of equations on the $k$ input packets, but solving the global linear system without considering the chunk structure usually has high computational cost. Therefore, we consider the following BP decoding of overlapped chunked codes.

The BP decoding includes multiple iterations. A chunk with transfer matrix $\mathbf{T}$ is said to be \emph{decodable} if $\mathbf{T}$ has full row rank. In the first iteration, all the decodable chunks are \emph{decoded} by solving \eqref{eq:decoding}, and the input packets involved in these decodable chunks are recovered. In each of the following iterations, undecoded chunks are first updated: Consider the updating of the $j$-th chunk. For each input packet in $\chunk_j$ that is decoded in the previous iteration, the value of this input packet is substituted into \eqref{eq:decoding}, reducing the number of unknown input packets in \eqref{eq:decoding}. If the updated \eqref{eq:decoding} becomes decodable for a chunk $j$, then decode the chunk and recover the input packets involved in the chunk. The BP decoding stops when no chunks become decodable in an iteration.
It is easily seen that the above decoding algorithm costs $\mathcal{O}(m^2+mL)$ finite field operations per packet.

\begin{definition}[Achievable rate]
We say a rate $R$ is \emph{achievable} by chunked codes using BP decoding if for any constant $\epsilon>0$, there exists a chunked code with $k\geq (R-\epsilon)mn$ input packets and $n$ chunks for all sufficiently large $n$ such that with probability at least $1-\epsilon$, when the BP decoding stops, at least $(R-\epsilon)mn$ input packets are recovered.
\end{definition}

\begin{remark}
  It is not necessary that the chunked code recovers all the input packets. When all the input packets are required to be recovered by the destination node, we can either retransmit the input packets that are not recovered, or use the precode technique as in Raptor codes~\cite{Shokrollahi}. 
\end{remark}

Our objective is to design an efficient class of overlapped chunked codes according to the rank distribution.
A natural upper bound on the achievable rates of chunked codes is established as follows.

\begin{proposition}\label{prop:upper}
  The achievable rate of chunked codes for transfer matrices with rank distribution $t=(t_0,t_1,\ldots,t_m)$ is upper bounded by $\bar{t}/m$, where
\begin{equation*}
  \bar{t}=\ex[\rank(\mathbf{T}_j)]=\sum_{i=1}^m it_i.
\end{equation*}
\end{proposition}
\begin{IEEEproof}
See Appendix~\ref{sec:proof-prop-refpr}.
\end{IEEEproof}

\section{Expander Chunked  Codes}
\label{Section:EC}

In this section, we introduce a family of overlapped chunked codes, named \emph{Expander Chunked (EC) codes}.\footnote{
EC codes were motivated by the expander graphs, and the expansion property was applied in the first analysis of EC codes to obtain a lower bound on the achievable rates\cite{tang12}. In this paper, we provide a better bound on the achievable rate without an explicit application of the expansion property, but the name of the code is preserved.}

\subsection{Code Description}

An EC code has three parameters:
 the number of chunks $n$, chunk size $m$ and degree $d$ ($3\leq d\leq m$). Let $k=n(m-d/2)$. Here, we assume $dn$ is even so that $k$ is an integer. An EC code is generated by a $d$-regular graph $G(V,E)$, called the \emph{generator (graph)}, where $V=\{1,2,\ldots,n\}$ is the node set and $E$ is the edge set. We will discuss the design of $G$ later in this paper.
The chunks in the EC code are constructed by the following steps.
\begin{enumerate}
  \item Label each edge $e\in E$ with a distinct integer in $\{1,\ldots,k\}$, and denote the integer by $i_e$. Label the rest $k-nd/2=(m-d)n$ integers in $\{1,\ldots,k\}$ evenly to the $n$ nodes in $V$, and denote the set of integers labelled to node $v$ by $\chunkset_v'$.

  \item Form $n$ chunks $\{\chunkset_v, 1\leq v \leq n\}$, where
      \begin{equation*}
        \chunkset_v=\chunkset_v' \cup \{i_e: e \text{ is incident to node } v\}.
      \end{equation*}
\end{enumerate}

Due to the one-to-one correspondence between nodes in $G$ and the chunks, we equate a node with its corresponding chunk henceforth in the discussion. We call $\chunkset_v$ chunk $v$, and $i_e$ an overlapping packet of chunk $v$.

As discussed in Section~\ref{sec:encoding-chunks}, EC code can be encoded causally. Specifically, the first step of the construction can be done as follows, where each index in $\{1,2,\ldots,k\}$ is used in an increasing order. First label node 1 with the first $m-d$ indices and label the $d$ edges incident to node 1 in an arbitrary order with the next $d$ indices. Then label node 2 with the next $m-d$ indices and label each of the edges incident to node 2 but unlabelled with a next index, and so on. Clearly, for any chunk $v$, the largest index in $\chunkset_v$ is less than or equal to $mv$.  See Fig.~\ref{fig:EC} for an illustration of this assignment of indices such that the chunks are suitable for causal encoding.

\begin{figure}[!t]
\centering

\begin{tikzpicture}[solid/.style={circle, draw,fill,minimum size=6pt,inner sep=0pt}, scale=1.2]

  \foreach \r/\l in {0/{1,2},1/{6,7},2/{10,11},3/{14,15},4/{17,18},5/{20,21}}
  {
    \node[solid, label=\r*60:\l] (n\r) at (60*\r:1.6cm) {};
  }
  \draw[thick] (n0) -- node[above, sloped] {3}
               (n1) -- node[above, sloped] {8}
               (n2) -- node[above, sloped] {12}
               (n3) -- node[below, sloped] {16}
               (n4) -- node[below, sloped] {19}
               (n5) -- node[below, sloped] {4} (n0);
  \draw[thick] (n0) -- node[pos=0.7,above, sloped] {5} (n4);
  \draw[thick] (n2) -- node[above, sloped] {13} (n5);
  \draw[thick] (n1) -- node[pos=0.3,below, sloped] {9} (n3);

   \node[right=of n0,scale=0.8, thick] {
    $\begin{aligned}
      \chunkset_1 & =\{1,2,3,4,5\} \\
      \chunkset_2 & =\{3,6,7,8,9\} \\
      \chunkset_3 & =\{8,10,11,12,13\} \\
      \chunkset_4 & =\{9,12,14,15,16\} \\
      \chunkset_5 & =\{5,16,17,18,19\} \\
      \chunkset_6 & =\{4,13,19,20,21\}
    \end{aligned}$
   };
\end{tikzpicture}
\caption{An EC code with $n=6$, $m=5$ and $d=3$. The generator graph of the code is a 3-regular graph with 6 nodes.}
\label{fig:EC}
\end{figure}

\subsection{Achievable Rates}

The performance of EC code with a particular generator graph is difficult to analyze. We instead analyze the performance of an EC code with a \emph{random $d$-regular graph} as the generator. There are various probability models for random $d$-regular graphs. We adopt the uniform model, \textit{i.e.}, $G$ is uniformly chosen from all $d$-regular graphs with node set $V$. One can obtain the similar result for the permutation model, the perfect matching model~\cite{ShiRRG}, etc.

The details of the performance analysis are provided in the next subsection, here we first characterize the achievable rates of EC codes under BP decoding. To state the main result, we need to introduce some notations. For any $3\leq d\leq m$, define a function $\alpha_d(y)$ over the interval $[0,1]$ as
\begin{equation}
\label{Eq: alpha}
\alpha_d(y)=\sum_{w=0}^{d-1}\binom{d-1}{w}y^w(1-y)^{d-1-w}\beta_w,
\end{equation}
where $\beta_w$ is defined in Theorem~\ref{lemma: iterative decoding}.
Note that
\begin{equation}
  \alpha_d(0)=\beta_0=t_m>0,
\end{equation}
and
\begin{equation}
\alpha_d(y)\leq 1,\quad y\in [0,1].
\end{equation}
We can further check that $\alpha_d(y)$ is monotonically increasing in $y$ (see Appendix~\ref{Appendix:Property}). With function $\alpha_d(y)$ and its functional powers, we introduce a sequence
\begin{equation}\label{eq:2}
  \alpha_d(0), \alpha_d^2(0),\alpha_d^3(0),\ldots,
\end{equation}
where $\alpha_d^{i+1}(0)=\alpha_d(\alpha_d^{i}(0))$ for all $i>0$. This sequence is well-defined since the range of $\alpha_d$ is in $[0,1]$. Further, since $\alpha_d(0) > 0$ and $\alpha_{d}(y)$ is monotonically increasing, we can check inductively that the sequence in \eqref{eq:2} is also monotonically increasing.
Since the sequence is bounded above, it must converge. Denote
\begin{equation*}
  \alpha_{d}^*=\lim_{i\rightarrow \infty} \alpha_{d}^{i}(0).
\end{equation*}
We further define
\begin{equation*}
  \tau_{d}=\alpha_{d+1}(\alpha_{d}^*),
\end{equation*}
and
\begin{equation*}
\lambda_d=1-(1-\alpha_d^*)^2.
\end{equation*}

\begin{theorem}
\label{Thm: EC}
EC codes with the degree $d$ and chunk size $m$ can achieve a rate at least $\tau_d(1-d/m)+\lambda_d d/(2m)$.
\end{theorem}

Note that, for any fixed degree $d$, the achievable rate given in Theorem~\ref{Thm: EC} is easy to calculate numerically.  So we can easily find a proper degree $d$ to maximize the achievable rate.

\subsection{Performance Analysis}
\label{sec:analysis}

We provide an analysis of the BP decoding of the EC code with a random $d$-regualr graph as the generator and prove Theorem~\ref{Thm: EC}.

\begin{definition}
  For any generator graph $G=(V,E)$, the \emph{$l$-neighborhood} of a node $v\in V$, denoted by $G_l(v)$, is the subgraph of $G$ induced by all the nodes $u$ with distance at most $l$ to $v$.
\end{definition}

After $l+1$ iterations of the BP decoding, whether all the input packets in chunk $v$ are recovered is determined by $G_l(v)$. Hence, we study the BP decoding performance $G_l(v)$.

\begin{definition}
 For any generator graph $G=(V,E)$, a node $v\in V$ is said to be $l$-decodable if $G_l(v)$ 
 all the input packets in chunk $v$ can be decoded when the decoding process is applied on $G_l(v)$.
\end{definition}

In the following, we set
\begin{equation*}
    l=\left \lfloor\frac{1}{3}\log_{d-1}n \right \rfloor.
\end{equation*}
We first show that a random regular graph has the \emph{locally tree-like} property, \textit{i.e.}, almost all the nodes in $G$ have their $l$-neighborhoods being trees.
\begin{lemma}
\label{Lemma: tree}
  For a random $d$-regular graph $G$ with $n$ nodes, let $T$ be the number of nodes with their $l$-neighborhoods being trees. Then, for any constant $\epsilon>0$,
  \begin{equation*}
    \Pr \{T>(1-\epsilon)n\} \geq 1-\mathcal{O}\left ( n^{-1/3}/\epsilon\right ).
  \end{equation*}
\end{lemma}
\begin{IEEEproof}
   Let $X_r$ be the number of cycles of length $r$ in $G$. One important fact is that a node whose $l$-neighborhood is not a tree must belong to a cycle with length less than or equal to $2l+1$. Therefore,
  \begin{equation}
  \label{Eq:relation}
    n-T\leq \sum_{r=3}^{2l+1}rX_r.
  \end{equation}
  Since $(d-1)^{2l+1}=o(n)$, it was shown in~\cite{McKayWW} that, for any $3\leq r\leq 2l+1$,
  \begin{eqnarray}
    \ex[X_r]&=&\frac{(d-1)^r}{2r}\left ( 1+ \mathcal{O}\left( \frac{r(r+d)}{n}\right)\right ) \nonumber \\
    &=&\frac{(d-1)^r}{2r}(1+o(1)). \label{Eq: numberofcycls}
  \end{eqnarray}
  Taking expectation on both sides of \eqref{Eq:relation} and substituting \eqref{Eq: numberofcycls} gives
  \begin{eqnarray*}
    \ex[n-T]&\leq &\sum_{r=3}^{2l+1}r\ex[X_r]\\
    &=&\sum_{r=3}^{2l+1}\frac{(d-1)^r}{2}(1+o(1))\\
    &=&\mathcal{O} \left ((d-1)^{2l+1}\right )\\
    &=&\mathcal{O} \left(n^{2/3}\right ).
   \end{eqnarray*}
  Finally, by Markov's inequality, we get
   \begin{eqnarray*}
    \Pr\{T\leq(1-\epsilon) n\}&=&\Pr \{n-T\geq \epsilon n\}\\
    &\leq & \frac{\ex[n-T]}{\epsilon n}\\
    &\leq & \mathcal{O}\left ( n^{-1/3}/\epsilon\right ).
  \end{eqnarray*}
\end{IEEEproof}

Now we show the probability that a node $v$ is $l$-decodable given that $G_l(v)$ is a tree. Note that the tree-based analysis of EC codes can be viewed as a variation of the and-or-tree analysis used for LT and LDPC codes.

\begin{lemma}
\label{Lemma: tree analysis}
Let $v\in V$ be a node such that $G_l(v)$ is a tree. Then for any constant $\epsilon>0$ and sufficiently large $n$,
\begin{itemize}
  \item the probability that chunk $v$ is $l$-decodable is at least $(1-\epsilon)\tau_{d}$, and
  \item the probability that an overlapping packet in chunk $v$ can be recovered by BP decoding on $G_l(v)$ is at least $(1-\epsilon)\lambda_d$.
\end{itemize}
\end{lemma}

\begin{IEEEproof}
  We first prove the first part. Consider the tree $G_l(v)$ rooted at $v$. Clearly, the root $v$ has $d$ children nodes and all other internal nodes have $d-1$ children nodes. Let $h_i$ be the probability that a node $u$ at level $i$ (here we assume that the node $v$ is at level $l$ and the leaves are at level $0$) is decodable when the decoding process of $u$ is restricted within the subtree of $G_l(v)$ rooted at $u$. In the following, we calculate $h_i$ in a bottom-up fashion.

For a leaf node $u$,  since it cannot get any help from other chunks in $G_l(v)$,
  \begin{equation*}
    h_0=\Pr\{\rank(\mathbf{T}_u)=m\}=t_m=\beta_0.
  \end{equation*}
  For any node $u$ at level $i$, $1\leq i\leq l-1$, suppose that $w$ out of the $d-1$ children nodes $v'$ of node $u$ are decodable when the decoding process of $v'$ is resticted within the subtree of $G_l(v)$ rooted at $v'$. Note that each of these children nodes (regarded as chunks) overlaps with chunk $u$ at a distinct packet. Therefore, when decoding $u$, these $w$ overlapping packets provide additional $w$ linearly independent coding vectors beyond $\mathbf{T}_u$. According to Theorem~\ref{lemma: iterative decoding}, the probability that $u$ is decodable is $\beta_w$. Since the local decoding processes of all the children nodes of node $u$ are mutually independent, we have
  \begin{eqnarray*} h_i&=&\sum_{w=0}^{d-1}\binom{d-1}{w}h_{i-1}^w(1-h_{i-1})^{d-1-w}\beta_w\\
   & =& \alpha_{d}(h_{i-1}).
  \end{eqnarray*}
  By induction, we have
  \begin{equation*}
    h_i=\alpha_{d}^{i+1}(0),\ i=0,1,\ldots,l-1.
  \end{equation*}
  Similarly, since the node $v$ in the level $l$ has $d$ children nodes,
  \begin{eqnarray*}
    h_l&=&\sum_{w=0}^{d}\binom{d}{w}h_{l-1}^w(1-h_{l-1})^{d-w}\beta_w\\
    &=& \alpha_{d+1}(h_{l-1})\\
    &=&\alpha_{d+1}(\alpha_{d}^{l}(0)).
  \end{eqnarray*}
  When $n\rightarrow \infty$, which implies $l\rightarrow \infty$, $\alpha_{d}^{l}(0)\rightarrow \alpha_{d}^*$. Therefore, $h_l\rightarrow \tau_{d}$ as $\alpha_{d+1}(y)$ is continuous. Hence, for any constant $\epsilon>0$,
  \begin{equation*}
    h_l>(1-\epsilon)\tau_d
  \end{equation*}
  for $n$ sufficiently large.

  Next we prove the second part. Let $u$ be an arbitrary children of node $v$. According to the above analysis, we know that node $u$ is decodable with probability $h_{l-1}=\alpha_d^l(0)$. Meanwhile, under the condition that chunk $u$ is not decodable, we can consider a new tree obtained by deleting the subtree rooted at $u$ from $G_l(v)$. Similarly, we can show that node $v$ can be decoded on the new tree with probability $\alpha_d(h_{l-1})=\alpha_d^{l+1}(0)$. Therefore, the common packet of chunk $u$ and chunk $v$ can be decoded with probability at least $1-(1-\alpha_d^l(0))(1-\alpha_d^{l+1}(0))$, which approaches $\lambda_d$ when $n$ goes to infinity. The proof is accomplished.
\end{IEEEproof}

Lemma~\ref{Lemma: tree} and Lemma~\ref{Lemma: tree analysis} together give a bound on the expected number of packets that can be recovered by BP decoding. Finally, we complete the proof of Theorem~\ref{Thm: EC} by showing that the number of recovered packets is sharply concentrated to its expectation.

\begin{IEEEproof}[Proof of Theorem~\ref{Thm: EC}]
Let $Z$ be the number of input packets recovered when the decoding process of every chunk is restricted within its $l$-neighborhoods, and let $T$ be the number of nodes whose $l$-neighborhood is a tree. According to Lemma~\ref{Lemma: tree analysis}, and noting that each chunk has $m-d$ non-overlapping packets and each of the $d$ overlapping packets only appear in two chunks, we have that for sufficiently large $n$,
  \begin{equation}
  \label{Eq: expectation}
    \ex[Z|T]\geq (1-\epsilon/4)(\tau_d(m-d)+\lambda_dd/2)T.
  \end{equation}

  Now consider an exposure martingale on $G$ as follows. Let
  \begin{equation}
  \label{Eq: Z0}
    Z_0=\mathbf{E}[Z|T],
  \end{equation}
  and for $i=1,2,\ldots,n$, let
  \begin{equation*}
    Z_i=\mathbf{E}[Z|\mathbf{T}_1,\mathbf{T}_2,\ldots,\mathbf{T}_i,T],
  \end{equation*}
  where $\mathbf{T}_i$ denotes the transfer matrix of chunk $\chunk_i$. The sequence $Z_0,Z_1,\ldots,Z_n$ gives a standard Doob martingale~\cite{Probability}. Recall that the decoding process of each node $v$ is restricted within the $l$-neighborhood $G_l(v)$. Therefore, the exposure of $\mathbf{T}_v$ affects the expected number of recovered packets by at most the number of nodes in $G_l(v)$ times the chunk size. More precisely, for each $1\leq i\leq n$,
  \begin{equation*}
    |Z_i-Z_{i-1}|\leq m|G_l(v)|=\Theta \left ((d-1)^l\right )=\Theta\left (n^{1/3}\right).
  \end{equation*}
  Applying the Azuma-Hoeffding Inequality~\cite{Probability}, we have
  \begin{eqnarray}
      \Pr\left\{Z_n\leq Z_0-\frac{\epsilon}{4} (\tau_d(m-d)+\lambda_dd/2)T\right\} &\leq & \exp\left(-\frac{\left(\frac{\epsilon}{4}(\tau_d(m-d)+\lambda_dd/2)T\right)^2}{2n\left (\Omega\left (n^{1/3}\right)\right)^2}\right) \nonumber \\
      & =& \exp\left(-\Omega\left (\epsilon^2n^{1/3}\right)\right).
      \label{Eq: Zn}
  \end{eqnarray}
  Combining (\ref{Eq: expectation}), (\ref{Eq: Z0}), (\ref{Eq: Zn}) and noting that $Z_n=Z$, we get
  \begin{equation}
  \label{Eq: EOC rate: Poisson case}
    \Pr\left\{Z\leq \left(1-\frac{\epsilon}{2}\right)(\tau_d(m-d)+\lambda_dd/2)T\right\}\leq \exp\left(-\Theta\left (\epsilon^2n^{1/3}\right)\right).
  \end{equation}
  Finally, since $T\geq (1-\epsilon/2)n$ almost surely according to Lemma~\ref{Lemma: tree}, and $Z$ is a natural lower bound on the number of packets that can be decoded by the BP decoding algorithm, we complete the proof of Theorem~\ref{Thm: EC}.
\end{IEEEproof}

\subsection{Generator Graph Design}

The above performance analysis implies that most $d$-regular graphs have the locally tree-like structure and hence the corresponding EC codes have the desired BP decoding performance. Therefore, the generator graph $G$ can be designed randomly. That is, we randomly generated a $d$-regular graph as the generator graph, which can be done in expected $\mathcal{O}(n)$ time by the McKay-Wormald algorithm~\cite{McKay}.  We will use this approach in our performance evaluation.

Since a randomly generated $d$-regular graph lacks a structure, we may need the whole adjacency matrix to preserve the graph. Note that the adjacency matrix is sparse and hence can be compresssed. Alternatively, we may just save the seed of the pseudorandom generator used for generating the $d$-regular graph.

Structured $d$-regular graphs can further simplify the generation and/or preservation of the EC code.  When $d=8$, Margulis' method~\cite{Margulis} gives a structured $8$-regular graph. However, currently we do not have an efficient algorithm for generating structured regular graphs with any parameters $d$ and $n$. Construction of structured regular graphs is of independent interest in mathematics and computer sciences, and many researches have been conducted on developing new approaches~\cite{Hoory}.

\section{Performance Evaluation}
In this section, we evaluate the performance of EC codes with comparison against the state-of-the-art overlapped chunked codes (H2T codes  \cite{HeidarzadehITW} and random annex codes (RAC)~\cite{LiTIT2011}) and coded chunked codes (BATS codes \cite{yang14} and L-chunked codes \cite{yang14ldpc}). For all evaluated chunked codes, we set $m=32$ and $q=256$, which give a good balance between the achievable rates and the encoding/decoding cost.

\subsection{Random Transfer Rank Distributions}

The performance of EC codes, as well as BATS codes and L-chunked codes, depend on the rank distribution $t=(t_0,t_1,\ldots,t_m)$. So we first evaluate the performance of EC codes for general rank distributions, which may provide some guidance on the application of EC codes.

Recall that the achievable rate of chunked codes is upper bounded by $\bar{t}/m$ (see Proposition~\ref{prop:upper}). For each fixed value $\bar{t}/m=0.5,0.6,0.7,0.8$, we sample a number of rank distributions\footnote{To the best of our knowledge, no efficient algorithms have been developed for uniformly sampling a rank distribution with a given mean value. Here, we use the following method for randomly sampling rank distributions. For a fixed $\bar{t}$, denote $a=\lfloor \bar{t}\rfloor$. We first sample a distribution $(t_0,t_1,\ldots,t_a)$ over the set $\{0,1,\ldots,a\}$ and a distribution $(t_{a+1},t_{a+2},\ldots,t_m)$  over the set $\{a+1,a+2,\ldots,,m\}$ using the method in~\cite{smith04}, which gives almost uniform sampling of distributions over the corresponding set. Let $\eta=(\sum_{i=a+1}^m it_i-\bar{t})/(\sum_{i=a+1}^m it_i-\sum_{i=0}^a it_i)>0$. Then we get a distribution  $(\eta t_0,\eta t_1,\ldots,\eta t_a, (1-\eta)t_{a+1},(1-\eta)t_{a+2},\ldots,(1-\eta)t_m)$, whose expectation is equal to $\bar{t}$.} and derive the corresponding achievable rates of EC codes, BATS codes and L-chunked codes numerically.
For EC codes, the achievable rate is given by Theorem~\ref{Thm: EC} with an optimized $d$. For BATS and L-chunked codes, the achievable rate is obtained by solving the corresponding degree distribution optimization problem.

The results are summarized in Table~\ref{EC:random}. From the table, we see that when $\bar t/m=0.5$, the average achievable rate of EC codes is much lower than the upper bound $0.5$. (Actually, EC codes perform worse when $\bar t/m$ is lower.) The reason is roughly as follows: each input packet in an EC code is duplicated as most once, so the total number of packets in an EC code $nm$ is no more than $2k$, where $k$ is the number of input packets. When $\bar t/m=0.5$, the effective number of received packets (removing the packets in each chunk that have linearly-dependent coefficient vectors) is about $nm/2 \leq k$. We see that EC codes in this case may not have enough redundancy for recovering a significant fraction of the input packets.

When the value of $\bar t/m$ becomes larger, the achievable rate of EC codes consistently becomes more close to $\bar t/m$. When $\bar t/m=0.8$, for example, the average achievable rate of EC codes is nearly $90$ percent of $\bar t/m$. It is not surprising to see that both BATS codes and L-chunked codes outperform EC codes due to the much more complicated encoding process and degree distribution optimization in the former codes.

By comparing the maximum and minimum achievable rates, we notice that the performance of EC codes varies significantly for different rank distributions, especially when $\bar t/m$ is relatively small. When $\bar t/m = 0.5$, for some rank distributions, EC codes achieve more than 80 percent of $\bar t/m$; while for some other rank distributions, EC codes can only achieve less than half of the rate of BATS/L-chunked codes.

\begin{table} \center
\caption{Achievable rates of EC/BATS/L-chunked codes with randomly sampled rank distributions of transfer matrices. For each value of $\bar t/m$, 10 rank distributions are sampled, and the minimum, maximum and average achievable rates of these ramples are given in the table.} \label{EC:random}
\begin{tabular}{cccc|ccc|ccc|ccc}
\toprule
\multirow{2}{*}{} &
\multicolumn{3}{c|}{$\bar{t}/m=0.5$} &
\multicolumn{3}{c|}{$\bar{t}/m=0.6$}&
\multicolumn{3}{c|}{$\bar{t}/m=0.7$} &
\multicolumn{3}{c}{$\bar{t}/m=0.8$}  \\
\cline{2-13}
  & avg. & min & max  & avg. & min & max  & avg. & min. & max.  & avg. & min. & max. \\
\midrule  \rowcolor[gray]{0.9}
EC & 0.294 & 0.184 & 0.411 & 0.523 & 0.508&0.532 & 0.591 & 0.569 & 0.619 & 0.719&0.694 &0.740\\
BATS & 0.497 & 0.495 & 0.498 &0.598 & 0.598&0.598 & 0.698 & 0.696 & 0.699 & 0.798 & 0.798 &0.759 \\
\rowcolor[gray]{0.9}
L-chunked & 0.478 & 0.470 & 0.486  & 0.581 & 0.570 & 0.592 & 0.687 & 0.673 & 0.697 &0.786 &0.778 &0.792\\
\bottomrule
\end{tabular}
\end{table}

In many potential applications of chunked codes, the rank distributions of the transfer matrices have certain features, instead of occurring purely randomly.  For instance, the number of packets in a chunk received by the destination node is a summation of multiple binomial random variables, which can be roughly approximated by a poisson random variable. Also, in an optimized transmission scheme, if the average packet loss rate over the network is higher, the number $M_j$ of packets transmitted for each chunk usually also becomes larger, so that the average rank $\bar{t}$ has a relatively large value~\cite{bin14}. In practice, EC codes can benefit from these features of rank distributions and achieve much higher rates than a rank distribution randomly generated. Therefore, in the remainder of this section, we focus on the performance of EC codes in a practical scenario.

\subsection{Line Networks}

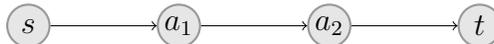
\begin{figure}
  \centering
  \begin{tikzpicture}[dot/.style={circle,draw=gray!80,fill=gray!20,thick,inner
 sep=1pt,minimum size=16pt}]
     \node[dot] (s) at(-2,0) {$s$};
     \node[dot] (a1) at(0,0) {$a_1$} edge[<-] (s);
     \node[dot] (a2) at(2,0) {$a_2$} edge[<-] (a1);
     \node[dot] (t) at(4,0) {$t$} edge[<-] (a2);
  \end{tikzpicture}
  \caption{Line network with length three. Node $s$ is the source node, node $t$
    is the destination node, and nodes
    $a_1$ and $a_2$ are the intermediate nodes that do not demand the input packets.}
  \label{fig:line}
\end{figure}

We consider a line network formed by tandem homogeneous links, each of which has the same packet loss probability $\epsilon$. Fig.~\ref{fig:line} illustrates a line network of length three. Line networks are generic building blocks of more complicated communication networks, and have attracted a lot of research interests \cite{pakzad05,niesen,vellambi11}. The chunk transmission schemes of line networks can be extended to general unicast networks and some multicast networks~\cite{bin14,yang14}, preserving the optimality.
In order to compare with the line network capacity directly, we instead evaluate the \emph{achievable network transmission rate}, i.e., the number of packets that are transmitted on average by one use of the network reliably.

We use the near-optimal chunk transmission scheme described in~\cite{bin14} over the line network. In this scheme, the chunks are transmitted in a sequential manner, and every node $v$, except for the destination node, transmits $M_j^{(v)}$ packets of each chunk $\chunk_j$, where $M_j^{(v)}$ is an integer-valued random variable. For the source node $s$, $M_j^{(s)}$ is just the variable $M_j$ defined in Section~\ref{sec: transmissionmodel}. For all the network nodes and chunks,
$M_j^{(s)}$ has the same mean value $\bar{M}$. For a fixed $\bar{M}$, the distribution of $M_j^{(v)}$ is optimized hop-by-hop according to the number of $j$-packets received/possessed by node $v$. The value of $\bar{M}$ is chosen such that $\bar{t}/\bar{M}$ is maximized, which is an upper bound on the network transmission rate that can be achieved by any chunked code under this transmission scheme.

We evaluate the performance of EC, BATS and L-chunked codes in line networks with different network lengths and packet loss probabilities. The results as well as some important parameters are summarized in Table~\ref{tab:1}-\ref{tab:3}. From these tables, we can see that when the network length or packet loss probability is larger, the optimized $\bar{M}$ is also larger, keeping $\bar{t}/\bar{M}$ at a high value, close to the network capacity (note that if the computational cost and/or buffer size of intermediate nodes is restricted to be $\mathcal{O}(1)$, the network capacity is smaller than $1-\epsilon$ and decreases when the network length grows~\cite{niesen,vellambi11}). Moreover, EC codes can achieve a network transmission rate that is about $91\%\sim 97\%$ of the bound $\bar{t}/\bar{M}$ and is about $80\%\sim 94.5\%$ of the network capacity $1-\epsilon$. This demonstrates the great real-world potential of EC codes.

\begin{table}
\caption{Achievable network transmission rates of chunked codes in line networks with $\epsilon=0.1$.}
     \label{tab:1}
        \centering
        \begin{filecontents*}{asrate1.txt}
    len         ub        EC      Gamma     BATS      M         LDPCCC
    2.0000    0.8789    0.8511    0.8538    0.8781   32.0000    0.8738 
    3.0000    0.8663    0.8249    0.8462    0.8656   33.0000    0.8523
    4.0000    0.8573    0.8171    0.8531    0.8565   33.0000    0.8531
    5.0000    0.8504    0.8091    0.8469    0.8496   33.0000    0.8469
    6.0000    0.8449    0.7954    0.8278    0.8441   34.0000    0.8399
\end{filecontents*}
\pgfplotstabletypeset[
        every even row/.style={
          before row={\rowcolor[gray]{0.9}}},
        every head row/.style={
          before row=\toprule,after row=\midrule},
        every last row/.style={
          after row=\bottomrule},
        columns={len,EC,LDPCCC,BATS,ub,M},
        columns/len/.style={
          column name=network length
          },
        columns/LDPCCC/.style={
          column name=L-chunked
          },
        columns/ub/.style={
          column name=$\bar t/ \bar M$
          },
        columns/M/.style={
          column name=$\bar M$
          },
        precision=3]
        {asrate1.txt}
\end{table}

\begin{table}
\caption{Achievable network transmission rates of chunked codes in line networks with $\epsilon=0.2$.}
     \label{tab:2}
        \centering
        \begin{filecontents*}{asrate2.txt}
    len         ub        EC      Gamma    BATS         M       LDPCCC
    2.0000    0.7728    0.7429    0.7531    0.7717   35.0000    0.7640
    3.0000    0.7571    0.7179    0.7516    0.7561   36.0000    0.7520
    4.0000    0.7462    0.7022    0.7393    0.7452   36.5000    0.7408
    5.0000    0.7380    0.6913    0.7310    0.7371   37.0000    0.7321
    6.0000    0.7313    0.6815    0.7258    0.7305   37.5000    0.7266

	
\end{filecontents*}
\pgfplotstabletypeset[
        every even row/.style={
          before row={\rowcolor[gray]{0.9}}},
        every head row/.style={
          before row=\toprule,after row=\midrule},
        every last row/.style={
          after row=\bottomrule},
        columns={len,EC,LDPCCC,BATS,ub,M},
        columns/len/.style={
          column name=network length
          },
        columns/LDPCCC/.style={
          column name=L-chunked
          },
        columns/ub/.style={
          column name=$\bar t/ \bar M$
          },
        columns/M/.style={
          column name=$\bar M$
          },
        precision=3]
        {asrate2.txt}
\end{table}
\begin{table}
\caption{Achievable network transmission rates of chunked codes in line networks with $\epsilon=0.4$.}
     \label{tab:3}
        \centering
        \begin{filecontents*}{asrate4.txt}
     len        ub      EC          Gamma    BATS       M       LDPCCC
    2.0000    0.5701    0.5327    0.5387    0.5688   44.0000    0.5585
    3.0000    0.5537    0.5230    0.5357    0.5526   46.0000    0.5429
    4.0000    0.5425    0.5044    0.5247    0.5417   48.0000    0.5388
    5.0000    0.5343    0.4933    0.5226    0.5336   49.0000    0.5226
    6.0000    0.5277    0.4842    0.5093    0.5271   50.0000    0.5232

	
\end{filecontents*}
\pgfplotstabletypeset[
        every even row/.style={
          before row={\rowcolor[gray]{0.9}}},
        every head row/.style={
          before row=\toprule,after row=\midrule},
        every last row/.style={
          after row=\bottomrule},
        columns={len,EC,LDPCCC,BATS,ub,M},
        columns/len/.style={
          column name=network length
          },
        columns/LDPCCC/.style={
          column name=L-chunked
          },
        columns/ub/.style={
          column name=$\bar t/ \bar M$
          },
        columns/M/.style={
          column name=$\bar M$
          },
        precision=3]
        {asrate4.txt}
\end{table}

\subsection{Comparison with Overlapped Chunked Codes}

\begin{figure}
  \centering
  \begin{tikzpicture}[scale=1]
    \begin{axis}[xmin=6000, xmax=14000, xlabel=num. of decodable input packets,no markers,
       ymin=0, ymax=1, legend pos=north west,ylabel=empirical CDF
      ]
       \addplot table {cdfEC.txt};
       \addlegendentry{EC}
       \addplot table {cdfRAC.txt};
       \addlegendentry{RAC}
       \addplot table {cdfH2T.txt};
       \addlegendentry{H2T}
    \end{axis}
  \end{tikzpicture}

  \caption{The empirical cumulative distribution function (CDF) of the number of decodable input packets. Here $m=32$, $n=500$, $q=2^8$ and $\bar{M}=36.5$.}
  \label{fig:cdf}
\end{figure}
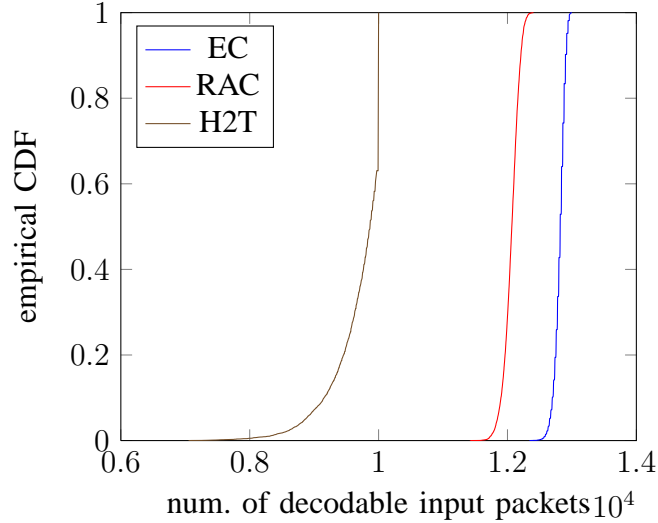

We then compare EC codes with two overlapped chunked codes: the chunked code with a head-to-tail type of overlapping (H2T) \cite{HeidarzadehITW} and random annex codes (RAC) \cite{LiTIT2011}.
Since we do not have the analytical results to calculate the achievable rates of these two codes,
we perform a simulation in a line network with length four and $\epsilon=0.2$ for the performance comparison.
For each code, we perform $10000$ runs of the simulation. In all the runs, the number of chunks in each code is set to be 500, which thus fixes the same transmission cost. The parameters involved in H2T and RAC are chosen optimally in the sense that the average number of decodable input packets is maximized. Note that given the parameters of a chunked code, the number of input packets is then determined, which varies over different classes of chunked codes.
The empirical cumulative distribution function of the number of decodable input packets for each code is plotted in Fig.~\ref{fig:cdf}. From this figure, we can see that EC codes outperform both H2T and RAC significantly.

\appendices
\section{Proof of Lemma~\ref{lemma: symmetric}}
\label{Appendix: transfer matrix}
 For matrix $\mathbf{A}$ and subspace $\mathsf{U}$, define
\begin{equation*}
  \mathbf{A}\mathsf{U}=\{\mathbf{A}\mathbf{z}:\mathbf{z}\in \mathsf{U}\}.
\end{equation*}
It can be checked that $\mathbf{A}\langle\mathbf{Z}\rangle=\langle\mathbf{A}\mathbf{Z}\rangle$.

Since $\mathsf{U}$ and $\mathsf{W}$ have the same dimension, there exists a full-rank $m\times m$ matrix $\mathbf{A}$ such that
  \begin{equation*}
    \mathsf{U}=\mathbf{A}\mathsf{W}.
  \end{equation*}
  Thus,
   \begin{equation}
  \label{eq:11}
   \Pr\{\langle\mathbf{T}_j\rangle=\mathsf{W}\}=\Pr\{\mathbf{A}\langle\mathbf{T}_j\rangle=\mathsf{U}\}
  =\Pr\{\langle\mathbf{A}\mathbf{T}_j\rangle=\mathsf{U}\}
  =\Pr\{\langle\mathbf{A}\mathbf{S}_j\mathbf{H}_j\rangle=\mathsf{U}\},
  \end{equation}
  where the first step follows by the invertibility of $\mathbf{A}$.

For any $s$ and $r$ such that $s\geq r$, denote the event $M_j=s$ and $\rank(\mathbf{S}_j)=r$ by $\mathcal{E}_{s,r}$, and define $\mathcal{S}_{s,r}$ to be the set of all $m\times s$ matrices with rank $r$. For any $S\in \mathcal{S}_{s,r}$, define $\mathcal{H}_S=\{H:\langle SH\rangle=\mathsf{U}\}$. Since $\mathbf{S}_j$ is totally random given $M_j=s$ and $\mathbf{A}$ is invertible, for any $S\in \mathcal{S}_{s,r}$,
\begin{equation*}
  \Pr\{\mathbf{S}=S|\mathcal{E}_{s,r}\}=\Pr\{\mathbf{A}\mathbf{S}=S|\mathcal{E}_{s,r}\}.
\end{equation*}

Using the assumption that $\mathbf{S}_j$ and $\mathbf{H}_j$ are independent conditioning on $M_j$ and $\rank(\mathbf{S}_j)$, we have
\begin{IEEEeqnarray*}{rCl}
\Pr\{\langle\mathbf{A}\mathbf{S}_j\mathbf{H}_j\rangle=\mathsf{U}\}
&=&\sum_{s,r:s\geq r} \Pr\{\langle\mathbf{A}\mathbf{S}_j\mathbf{H}_j\rangle=\mathsf{U} | \mathcal{E}_{s,r}\}\Pr\{\mathcal{E}_{s,r}\}\\
&=& \sum_{s,r:s\geq r}\sum_{S\in \mathcal{S}_{s,r}}\sum_{H\in \mathcal{H}_S}\Pr\{\mathbf{A}\mathbf{S}_j=S, \mathbf{H}_j=H|\mathcal{E}_{s,r}\}\Pr\{\mathcal{E}_{s,r}\}\\
&=& \sum_{s,r:s\geq r}\sum_{S\in \mathcal{S}_{s,r}}\sum_{H\in \mathcal{H}_S}\Pr\{\mathbf{A}\mathbf{S}_j=S|\mathcal{E}_{s,r}\}\Pr\{ \mathbf{H}_j=H|\mathcal{E}_{s,r}\}\Pr\{\mathcal{E}_{s,r}\}\\
&=& \sum_{s,r:s\geq r}\sum_{S\in \mathcal{S}_{s,r}}\sum_{H\in \mathcal{H}_S}\Pr\{\mathbf{S}_j=S|\mathcal{E}_{s,r}\}\Pr\{ \mathbf{H}_j=H|\mathcal{E}_{s,r}\}\Pr\{\mathcal{E}_{s,r}\}\\
&=&\sum_{s,r:s\geq r}\sum_{S\in \mathcal{S}_{s,r}}\sum_{H\in \mathcal{H}_S}\Pr\{\mathbf{S}_j=S, \mathbf{H}_j=H|\mathcal{E}_{s,r}\}\Pr\{\mathcal{E}_{s,r}\}\\
&=& \sum_{s,r:s\geq r} \Pr\{\langle\mathbf{S}_j\mathbf{H}_j\rangle=\mathsf{U} | \mathcal{E}_{s,r}\}\Pr\{\mathcal{E}_{s,r}\}\\
&=& \Pr\{\langle\mathbf{S}_j\mathbf{H}_j\rangle=\mathsf{U}\}\\
&=&\Pr\{\langle \mathbf{T}_j\rangle=\mathsf{U}\}.
\end{IEEEeqnarray*}
The proof is completed by combining the above equality with \eqref{eq:11}.

\section{Proof of Theorem~\ref{lemma: iterative decoding}}
\label{sec:iterativedecoding}

The proof is based on the uniformity property of transfer matrices given in Lemma~\ref{lemma: symmetric} together with counting. For a subspace $\mathsf{U}$, denote its dimension by $\text{dim}(\mathsf{U})$.  Since
  \begin{equation*}
    \rank([\mathbf{T}_j\text{ }\mathbf{D}])=\rank(\mathbf{T}_j)+\rank(\mathbf{D})-\text{dim}(\langle \mathbf{T}_j\rangle \cap \langle \mathbf{D} \rangle),
  \end{equation*}
  we have that, for any $m-w\leq i\leq m$,
  \begin{equation*}
    \Pr \{\rank([\mathbf{T}_j\text{ }\mathbf{D}])=m|\rank(\mathbf{T}_j)=i\}=\Pr\{\text{dim}(\langle \mathbf{T}_j\rangle \cap \langle \mathbf{D}\rangle)=i+w-m|\text{dim}(\langle\mathbf{T}_j\rangle)=i\}.
  \end{equation*}
  As there are ${m \brack i}$  $i$-dimensional subspaces of $\mathbb{F}_q^m$, and
  $q^{(m-i)(m-w)}{w\brack m-i}$ $i$-dimensional subspaces of $\mathbb{F}_q^m$ such that $\text{dim}(\langle \mathbf{T}_j\rangle \cap \langle \mathbf{D}\rangle)=r+w-i$ (ref.~\cite{andrews1998theory, gadouleau2010packing}), by Lemma~\ref{lemma: symmetric}, we have
  \begin{equation*}
  \Pr\{\text{dim}(\langle \mathbf{T}_j\rangle \cap \langle \mathbf{D}\rangle)=i+w-m|\text{dim}(\langle\mathbf{T}_j\rangle)=i\}=\frac{q^{(m-i)(m-w)}{w\brack m-i}}{{m \brack i}}.
  \end{equation*}
  Therefore,
  \begin{align*}
    \Pr \{\rank([\mathbf{T}_j\text{ }\mathbf{D}])=m\}=&\sum_{i=0}^m \Pr \{\rank([\mathbf{T}_j\text{ }\mathbf{D}])=m|\rank(\mathbf{T}_j)=i\}\Pr\{\rank(\mathbf{T}_j)=i\}\\
    =& \sum_{i=m-w}^{m}\frac{q^{(m-i)(m-w)}{w\brack m-i}}{{m \brack i}}t_i.
  \end{align*}

\section{Proof of Proposition~\ref{prop:upper}}
\label{sec:proof-prop-refpr}

  Assume that $\lambda=\bar{t}/m+\delta$, $\delta>0$ is achievable by chunked codes. Fix $\epsilon = \delta/2$, by the definition of achievable rates,
there exists a chunked code with $n$ chunks for all sufficiently large $n$ such that at least $(\lambda-\epsilon)mn$ input packets are recovered with probability at least $1-\epsilon$.

Note that in the decoding of a chunked code, only received packets of a chunk with linearly independent coefficient vectors are useful. Therefore, the number of decodable input packets is upper bounded by $\sum_{j=1}^{n} \rank(\mathbf{T}_j)$. Then we have the decoding error probability
  \begin{align*}
   P_{\text{err}} \geq & \Pr\left\{\sum_{j=1}^{n} \rank (\mathbf{T}_j)<(\lambda-\epsilon) m n\right \}\\
    =& \Pr\left \{\sum_{j=1}^{n} \rank (\mathbf{T}_j)<(\bar{t}+m\delta/2)n\right\}\\
    \geq & 1-e^{-\frac{ m^2\delta^2 n}{12\bar{t}}},
  \end{align*}
  where the last inequality follows from the Chernoff bound. For a sufficiently large $n$, we have $P_{\text{err}} > \epsilon$, a contradiction!

\section{Property of Function $\alpha_d(y)$}
\label{Appendix:Property}

\begin{lemma}\label{lemma:alpha}
  For $0\leq \beta_0 \leq \beta_1\leq \cdots \leq \beta_d \leq 1$,
  the function $\alpha_d(y)=\sum_{w=0}^{d-1}\binom{d}{w}y^w(1-y)^{d-1-w}\beta_w$ has range $[\beta_0,1]$ and is monotonically increasing in $y$.
\end{lemma}
\begin{IEEEproof}
Since $\beta_w\leq 1$ and $\beta_w$ is monotonically increasing in $w$,
    \begin{equation}
      \alpha_d(y)\leq\sum_{w=0}^{d-1}\binom{d-1}{w}y^w(1-y)^{d-1-w}=1.
    \end{equation}
It is clear that $\alpha_d(0)=\beta_0$, so it is sufficient to show that $\alpha_d(y)$ is monotonically increasing. The derivative of $\alpha_d(y)$ is
    \begin{IEEEeqnarray*}{rCl}
        (\alpha_d(y))^\prime & = & \sum_{w=0}^{d-1}\binom{d-1}{w}\left (wy^{w-1}(1-y)^{d-1-w}-(d-1-w)y^w(1-y)^{d-2-w}\right)\beta_w\\
        & = & \IEEEyesnumber     \label{Eq: increasing}
        \sum_{w=1}^{d-1}\binom{d-1}{w} wy^{w-1}(1-y)^{d-1-w}\beta_w
        -\sum_{w=0}^{d-2}\binom{d-1}{w}(d-1-w)y^w(1-y)^{d-2-w}\beta_w.
    \end{IEEEeqnarray*}
    The second term of the right hand side in (\ref{Eq: increasing}) can be transformed into
    \begin{equation*}
      \sum_{w=1}^{d-1}\binom{d-1}{w-1}(d-w)y^{w-1}(1-y)^{d-1-w}\beta_{w-1}
      =\sum_{w=1}^{d-1}\binom{d-1}{w}wy^{w-1}(1-y)^{d-1-w}\beta_{w-1}.
    \end{equation*}
    Therefore,
    \begin{equation*}
        (\alpha_{d}(y))^\prime=\sum_{w=1}^{d-1}\binom{d-1}{w}wy^{w-1}(1-y)^{d-1-w}(\beta_w-\beta_{w-1})\geq 0,
    \end{equation*}
    since $\beta_{w}\geq \beta_{w-1}$. The proof is accomplished.
\end{IEEEproof}



\ifCLASSOPTIONcaptionsoff
  \newpage
\fi



%
%

\bibliographystyle{IEEEtranTCOM}
\bibliography{./IEEEabrv,./eoc}

%




\end{document}